\documentclass[pra,aip,reprint]{revtex4-1}

\usepackage{graphicx,amsmath,braket, cancel}
\usepackage{appendix}

\begin{document}

\title{Toward improved quantum simulations and sensing with trapped two-dimensional ion crystals via parametric amplification}

\author{M.~Affolter}
\affiliation{National Institute of Standards and Technology, Boulder, Colorado 80305, USA}
\author{W.~Ge}
\affiliation{Department of Physics, University of Rhode Island, Kingston, Rhode Island 02881, USA}
\affiliation{Department of Physics, Southern Illinois University, Carbondale, Illinois 62901, USA}
\author{B.~Bullock}
\affiliation{National Institute of Standards and Technology, Boulder, Colorado 80305, USA}
\affiliation{Department of Physics, University of Colorado, Boulder, Colorado 80309, USA}
\author{S.~C.~Burd}
\affiliation{Department of Physics, Stanford University, Stanford, California 94305, USA}
\author{K.~A.~Gilmore}
\affiliation{National Institute of Standards and Technology, Boulder, Colorado 80305, USA}
\affiliation{Department of Physics, University of Colorado, Boulder, Colorado 80309, USA}
\author{J.~F.~Lilieholm}
\affiliation{National Institute of Standards and Technology, Boulder, Colorado 80305, USA}
\affiliation{Department of Physics, University of Colorado Boulder, Boulder, Colorado 80309, USA}
\author{A.~L.~Carter}
\affiliation{National Institute of Standards and Technology, Boulder, Colorado 80305, USA}
\author{J.~J.~Bollinger}
\affiliation{National Institute of Standards and Technology, Boulder, Colorado 80305, USA}

\date{\today}

\begin{abstract}
Improving coherence is a fundamental challenge in quantum simulation and sensing experiments with trapped ions. Here we discuss, experimentally demonstrate, and estimate the potential impacts of two different protocols that enhance, through motional parametric excitation, the coherent spin-motion coupling of ions obtained with a spin-dependent force. The experiments are performed on 2D crystal arrays of approximately 100 $^9$Be$^+$ ions confined in a Penning trap. By modulating the trapping potential at close to twice the center-of-mass mode frequency, we squeeze the motional mode and enhance the spin-motion coupling while maintaining spin coherence. With a stroboscopic protocol, we measure $5.4 \pm 0.9$ dB of motional squeezing below the ground-state motion, from which theory predicts a $10$-dB enhancement in the sensitivity for measuring small displacements using a recently demonstrated protocol [K. A. Gilmore \textit{et al}., Science \textbf{373}, 673 (2021)].  With a continuous squeezing protocol, we measure and accurately calibrate the parametric coupling strength.  Theory suggests this protocol can be used to improve quantum spin squeezing, limited in our system by off-resonant light scatter.  We illustrate numerically the trade-offs between strong parametric amplification and motional dephasing in the form of center-of-mass frequency fluctuations for improving quantum spin squeezing in our set-up.

\end{abstract}


\maketitle

\section{Introduction}
\label{Intro}

Trapped-ion systems have demonstrated high-fidelity quantum logic gates \cite{BallanceHF,GaeblerHF,MckayHF,SrinivasHF,SawyerHF}, spin squeezing \cite{BohnetApp,MeyerSS,PezzeReview,CoxSS,HostenSS}, the generation of many-ion entangled states \cite{FriisEntangled,LinkeEntangled}, and motional sensing below the ground-state motion of the ions \cite{GilmoreScience,AffolterPhase,GilmoreSensing,BurdPara}.  In these experiments, the interaction between qubits is engineered by coupling the spins to the collective motion of the ions using lasers or magnetic-field gradients.  Depending on the source of decoherence, stronger spin-motion coupling can enable higher-fidelity gates and improved quantum sensing.  However, the interaction strength is limited by the available laser power or current that can be driven through the trap electrodes.  In the case of laser-based interactions, the dominant source of decoherence is typically spin decoherence from spontaneous emission, so increasing the laser power alone will not necessarily improve the fidelity.

A viable approach to stronger spin-motion coupling that overcomes some of these technical and fundamental challenges is parametric amplification \cite{GePara,GePRA}.  By modulating the trapping potential at twice the motional mode frequency, one quadrature of motion is amplified while the orthogonal quadrature is attenuated \cite{HeinzenPara}.  This leads to motional squeezing and an amplification of the spin-motion coupling strength.  Improved displacement sensing of a single ion \cite{BurdPara} and faster two-qubit ion gates have recently been achieved with this technique \cite{BurdGate}.  Here we experimentally characterize both a stroboscopic and a continuous squeezing protocol on two-dimensional  (2D) crystal arrays of approximately $100$ ions.  Based on this characterization we estimate the potential improvements for displacement sensing and spin squeezing with large trapped-ion crystals.

In the stroboscopic protocol, parametric amplification is applied over a discrete interval to create a squeezed motional state.  The size of this motional squeezing is experimentally characterized through a phase-coherent Ramsey sequence to measure the variance of the center-of-mass (c.m.) motion in the squeezed and unsqueezed directions.  The c.m.\ motion is squeezed by $5.4\pm0.9$ dB $  (10.8\pm1.8$ dB) below the ground-state motional uncertainty (variance).  By applying this level of squeezing to amplify a spin-independent displacement, theory predicts a $10$-dB improvement in the sensitivity for measuring small displacements, accounting for c.m.\ frequency fluctuations. This has the potential to improve the sensitivity for measuring small displacements obtained with the protocol demonstrated in Ref. \citenum{GilmoreScience} from 8.8 dB to nearly 19 dB below the standard quantum limit (SQL). Here the SQL is given by the ground-state c.m.\ mode zero-point fluctuations.

For faster higher-fidelity quantum simulations, we also investigate a continuous squeezing protocol \cite{BurdGate}.  In this experiment, the parametric drive is applied simultaneously with the spin-dependent force to amplify the strength of the spin-motion coupling.  Theory \cite{GePara} shows that this protocol leads to an equivalent interaction Hamiltonian with a rescaled detuning and stronger interaction strength.  By measuring this rescaled detuning, we experimentally determine the parametric coupling strength $g$, which agrees well with predictions from a numerical model.  As an application of this continuous protocol, we explore theoretically how this increased interaction strength will lead to an improvement in quantum spin squeezing \cite{BohnetApp}.  Our model suggests that frequency fluctuations of the c.m.\ mode will be the primary limitation to the enhancement.  With a reduction of the frequency fluctuations from our current 40 Hz to 10 Hz, theory predicts 14 dB of spin squeezing is possible for a 400-ion crystal.

The rest of the paper is structured as follows. In Sec.~\ref{ExpApp} we describe the Penning trap setup and the implementation of parametric amplification.  In Sec.~\ref{Strobo} the stroboscopic squeezing protocol is discussed.  We describe how the motional squeezing is characterized and the predicted improvements in the sensitivity of detecting small displacements.  Section~\ref{ConProto} focuses on the continuous squeezing protocol.  Experimental data with and without continuous parametric amplification are presented, from which the parametric coupling strength is extracted without precise control of the phase of the parametric drive.  Theory predicts that this continuous parametric amplification protocol will enable improved spin squeezing in the presence of decoherence due to off-resonant light scatter \cite{BohnetApp}. The amount of improvement will likely be limited by the frequency stability of the c.m.\ mode. A summary is given in Sec. V.

\section{Experimental Apparatus}
\label{ExpApp}
\begin{figure}
\includegraphics[width=0.46\textwidth]{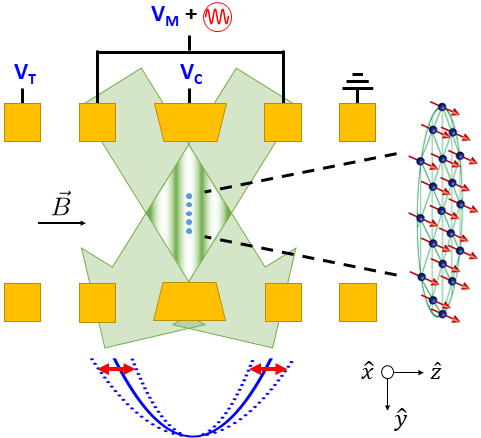}%
\caption{Simplified cross-sectional schematic of the Penning trap.  To generate the $\omega_{z}/2\pi=1.59$ MHz axial confinement potential (blue solid curve), voltages $V_{C}=-2$ kV and $V_{M}=-1.75$ kV are applied to three of the ring electrodes (yellow) with radius $R_{W}=1$ cm.  This confining potential is modulated to induce parametric amplification by applying a sinusoidal voltage of zero-to-peak amplitude $V_p$ on top of $V_{M}$.  A small tuning voltage $V_{T}=-43$\,V shifts the ion crystal (blue dots) to the null of this modulated potential.  These electrodes are held in the room-temperature bore of a $B=4.5\,$T superconducting magnet to provide radial confinement of the ions.  When cooled, the ions form a 2D crystal with an ion-ion spacing of approximately 15 $\mu$m, resulting in a crystal with a radius of approximately 100 $\mu$m.  The two optical dipole force beams (green) intersect at a $20^{\circ}$ angle at the ions to form the 1D traveling wave potential with a wavelength of 900 nm and a frequency $\mu$ that is equal to the frequency difference between the two beams.}
\label{TrapGeo}
\end{figure}

In our experimental apparatus \cite{AffolterPhase}, 2D crystal arrays of approximately $100$ $^{9}$Be$^{+}$ ions are confined in a Penning trap as shown in the simplified schematic of Fig.~\ref{TrapGeo}.  The ions are confined axially by a quadratic potential formed by static voltages applied to a stack of cylindrical electrodes. Radial confinement results from $\vec{E}\times \vec{B}$ induced rotation of the crystal. This rotation of the crystal through the $\vec{B} = 4.5\hat{z}$ T magnetic field ($^{9}$Be$^{+}$ cyclotron frequency of $\Omega_{c}/=7.6\,$MHz) produces a radially confining Lorentz force.  In the rotating frame of the crystal, to a very good approximation the confining potential can be written as
\begin{equation} \label{TrapPot}
    q\phi_{\text{trap}} = \frac{1}{2}M\omega_{z}^{2}\left(z^{2}+\beta_{r}\rho^{2}\right),
\end{equation}
where $M$ is the mass and $q$ is the charge of $^9$Be$^+$, $z$ $(\rho)$ is the axial distance (cylindrical radius) from the trap center, and the axial c.m.\ oscillation frequency $\omega_{z}/=1.59$ MHz quantifies the strength of this confining potential.  The magnetic-field strength and rotation frequency determine the relative strength of the radial confinement
\begin{equation} \label{RadConf}
   \beta_{r} = \frac{\omega_{r}\left(\Omega_{c}-\omega_{r}\right)}{\omega_{z}^{2}}-\frac{1}{2},
\end{equation}
which we control by setting the rotation frequency $\omega_{r}/=180$ kHz through the use of a weak dipole rotating wall potential \cite{HuangRW} [neglected in Eq.~\eqref{TrapPot}].

To parametrically amplify the axial c.m.\ mode, the confining potential is modulated at near twice the mode frequency.  This modulation is achieved by applying a sinusoidal voltage with an amplitude ranging from $1$~V to $51$~V at a frequency $\omega_{p}\sim2\omega_{z}$ in addition to the confining voltage $V_{M}=-1.75$\,kV (see Fig.~\ref{TrapGeo}).  Ideally, this modulation parametrically amplifies the c.m.\ motion along one quadrature and attenuates the motion in the orthogonal quadrature, producing motional squeezing.

However, if the ion crystal is not in the null of the modulated potential, motion is also directly driven at $\omega_{p}$.  To minimize this motion, we modulate the confining potential at an off-resonant frequency $\omega_{p}/\sim1.7$\,MHz, and detect this driven motion using techniques similar to those in Ref.~\citenum{GilmoreSensing}.  A small tuning voltage $V_{T}$ is then applied to one of the end cap electrodes, which shifts the crystal axially, and this experiment is repeated.  Through this process, the axial position with the minimum driven motion (null of the modulated potential) is determined to within $0.1\,\mu$m.

For our experiments, the two qubit states are the $\ket{\uparrow}\equiv|S_{1/2},m_{s}=+1/2\rangle$ and $\ket{\downarrow}\equiv|S_{1/2},m_{s}=-1/2\rangle$ valence electron spin projections of the $^{2}S_{1/2}$ electronic ground state, which are separated by approximately $124\,$GHz.  Global spin rotations are implemented with a microwave source with a $\pi-$pulse duration $t_{\pi}$ of about 50~$\mu s$ \cite{Britton_fluct}.

In this apparatus axial motion is coupled to the spins through the implementation of a spin-dependent optical dipole force (ODF).  Two far-detuned approximately $313\,$nm beams are overlapped at a $20^{\circ}$ angle at the ions, as shown in Fig.~\ref{TrapGeo}, to form a moving 1D optical lattice with an effective wavelength of $900\,$nm and a tunable beat frequency $\mu$.  As described in the Supplementary Information of Ref.~\citenum{BrittonQSim}, the frequency and polarizations of the ODF laser beams are adjusted to null the ac Stark shift of each beam and to produce a force on the $\ket{\uparrow}$ state that is equal and opposite to the force on the $\ket{\downarrow}$ state.

With the assumption that the ODF couples only to the c.m.\ mode, the interaction Hamiltonian describing this spin-motion coupling is \cite{GePara,GePRA}
\begin{equation} \label{HODF}
    \hat{H}_{\mathrm{ODF}}=\frac{\hbar f}{2\sqrt{N}}\left(\hat{a}e^{-i\phi_{\text{ODF}}}+\hat{a}^{\dagger}e^{i\phi_{\text{ODF}}}\right)\sum_{i}^{N}\hat{\sigma}^z_i-\hbar\delta\hat{a}^{\dagger}\hat{a},
\end{equation}
where $f$ represents the strength of the ODF interaction, $\hat{\sigma}^z_i$ is the Pauli $z$ operator for spin $i$, $\delta=\mu-\omega_{z}$ is the detuning of the ODF beat note frequency $\mu$ from the c.m.\ mode, $\hat{a}^{\dagger} (\hat{a})$ is the raising (lowering) operator for the c.m.\ mode, and $\phi_{\text{ODF}}$ is the phase of the optical dipole force. When the ODF is applied for a duration $\tau$ on-resonance $\delta=0$, a spin-dependent displacement $\hat{D}_{\text{sd}}(\alpha)\equiv\exp[(\alpha \hat{a}^{\dagger}-\alpha^\ast \hat{a})\sum \hat{\sigma}^z_i]$ is created with $\alpha=-i f e^{i\phi_{\text{ODF}}}\tau/(2\sqrt{N})$.

\section{Stroboscopic Protocol}
\label{Strobo}
\subsection{Characterizing Motional Squeezing}

\begin{figure}
\includegraphics[width=0.46\textwidth]{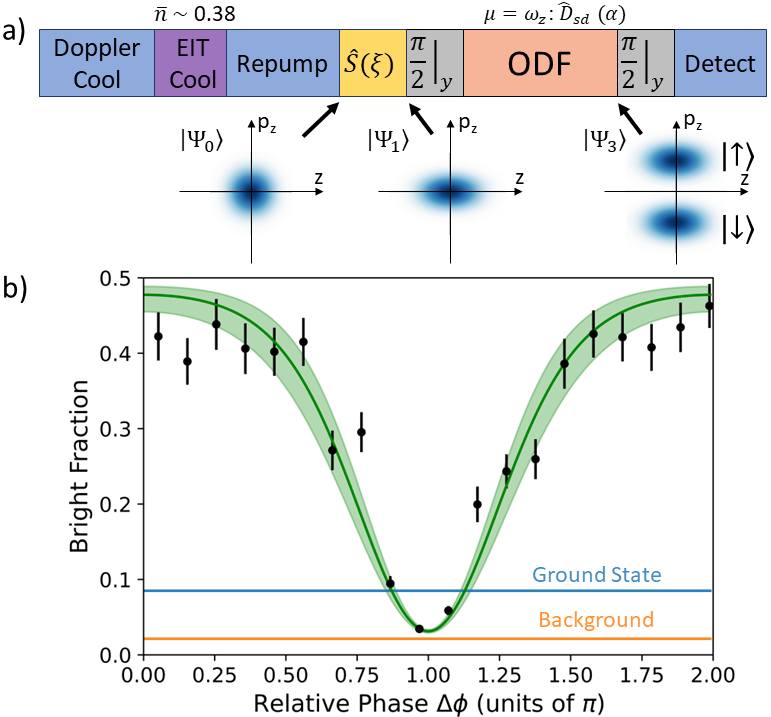}%
\caption{(a) Motional squeezing resulting from parametric amplification is characterized by entangling the spin and motion of the ions through a spin-dependent displacement and then measuring the coherence of the spins, which is sensitive to the overlap of the displaced, squeezed states.  The relative phase between the parametric drive (yellow block), which squeezes the motion, and the spin-dependent displacement (orange block) is controlled.  To reduce spin decoherence from magnetic-field fluctuations, the spin-dependent displacement is performed in a spin-echo protocol where $\phi_{\text{ODF}}$ in the second arm is advanced by $\pi$.  This is equivalent to the Ramsey sequence shown in the block diagram.  The phase-space diagram shows the c.m.\ motional mode in a frame rotating at $\omega_z$. Axis labels represent c.m.\ momentum [$p_z\propto$ Im$(\alpha)$] and position [$z\propto $ Re$(\alpha)$] in the c.m.\ mode rotating frame.  (b) The measured bright fraction (black points) depends on the relative phase $\Delta\phi$ [see Eq.~\eqref{stroboscopic deltaphi}], shown with a theoretical fit (green curve, Eq.~\eqref{BrightFrac}). Confidence intervals (green shaded region) represent uncertainty in the fitted $\bar{n}_z$. When the displacement is parallel (perpendicular) to the squeezed axis, corresponding to $\Delta\phi =0$ ($\Delta\phi = \pi$), the overlap of the displaced motional states is small (large) and the resulting bright fraction is high (low). The fitted value for the reduced motional uncertainty of the squeezed state is $5.4 \pm 0.9$ dB below the motional uncertainty of the ground state. The blue line is the signal that would be obtained from the c.m.\ motional ground state without squeezing, calculated from Eq.~\eqref{BrightFrac} with $r,\bar{n}_z=0$. Error bars represent the standard deviation in the bright fraction over the 50 experimental trials.}
\label{PhaseProto}
\end{figure}

To characterize the motional squeezing from parametric amplification, we use a stroboscopic protocol as shown in Fig.~\ref{PhaseProto}(a).  First, pulses of Doppler and electromagnetically induced transparency (EIT) cooling are sequentially applied to cool the c.m.\ mode to $\bar{n}_{z}=0.38\pm0.2$, and the other axial drumhead modes to near their motional ground state~\cite{JordanEIT, AthreyaEIT}.  A repump laser then initializes the spins in the state $\ket{\uparrow}^{\otimes N}$. The c.m.\ motion is then squeezed, followed by an analysis of the motional state with a resonant $(\mu=\omega_z)$ spin-dependent force \cite{HomeSqueeze}.  

For initial spin states that are product states and resonant applications of the spin-dependent force, the spins remain uncorrelated and the calculation of the expectation value of the composite spin of the system reduces to calculating the expectation value of an individual spin \cite{SawyerApp}.  We also begin by treating the c.m.\ motional state as a Fock state and extend this treatment to a thermal state by averaging a mixture of Fock states.   Therefore, after the spins are optically pumped, we assume the system is in the state $\ket{\Psi_{0}}=\ket{\uparrow}\ket{n}$, where $\ket{n}$ is the harmonic oscillator Fock state of the c.m.\ mode. 

The c.m.\ mode is squeezed by briefly modulating the trap potential to induce parametric amplification, which is described by the interaction Hamiltonian \cite{GePara,GePRA}
\begin{equation} \label{HMoSqu}
    \hat{H}_{\mathrm{s}}=i\hbar\frac{g}{2}(\hat{a}^{2}e^{-i\theta}-\hat{a}^{\dagger2}e^{i\theta}),
\end{equation}
where $g$ and $\theta$ are the parametric coupling strength and phase, respectively.  The parametric coupling strength is dependent on the amplitude of the voltage modulation $V_{p}$ of the quadratic trapping potential.  By numerically modeling the trap geometry and applied voltages, we calculate the expected $g$ for an applied modulation amplitude $V_{p}$.  

Application of this Hamiltonian for duration $t_{s}$ implements the unitary squeezing operator
\begin{equation} \label{SqOp}
    \hat{S}(\xi)\equiv \exp\left[\frac{1}{2}\left(\xi^{*}\hat{a}^{2}-\xi\hat{a}^{\dagger2}\right)\right],
\end{equation}
where $\xi(r,\theta)=r \exp(i\theta)$ and $r=gt_{s}$.  Along the squeezed axis the motional uncertainty is reduced by $\exp(-r)$ and in the orthogonal direction the motional uncertainty is amplified by $\exp(r)$, so the phase-space uncertainty area is preserved.  For this sequence, the modulation at $\omega_{p}=2\omega_{z}\sim2\pi\times3.18$\,MHz is typically applied for a duration of about $t_{s}=40\,\mu$s, which squeezes the initial state into the squeezed motional state $\ket{\Psi_{1}}=\ket{\uparrow} \hat{S}(\xi)\ket{n}$.  A phase-space sketch of this squeezed state is shown in Fig.~\ref{PhaseProto}(a).

We define the relative phase of the parametric drive to the spin-dependent displacement as  
\begin{equation} \label{stroboscopic deltaphi}
    \Delta \phi = \theta-2\phi_{\text{ODF}}-\pi.
\end{equation}
The factor of 2 is due to the parametric drive occurring at twice the frequency of the ODF beatnote, and the additional $\pi$ is to define in phase ($\Delta\phi=0$) as when the squeezed axis and spin-dependent displacements are aligned in the same direction.  This relative phase $\Delta \phi$ is actively stabilized and controlled \cite{AffolterPhase}.

The amount of motional squeezing is experimentally determined by measuring, through a Ramsey sequence, the coherence of the spins under the application of a spin-dependent ODF as a function of the relative phase $\Delta\phi$ \cite{HomeSqueeze}. A first microwave $\pi/2$-pulse rotates the spins to create  
\begin{equation} \label{FirstRot}
    \ket{\Psi_{2}}=\hat{R}\left(\frac{\pi}{2},0\right)\ket{\Psi_{1}}=\frac{\ket{\uparrow}+\ket{\downarrow}}{\sqrt{2}}\hat{S}(\xi)\ket{n},
\end{equation}
where we define the following qubit rotation matrix:
\begin{equation}\label{RotationMatrix}
    \hat{R}\left(\theta_r,\phi_r\right)=
    \begin{pmatrix}
        \cos(\frac{\theta_r}{2}) & -e^{-i\phi_r}\sin(\frac{\theta_r}{2})\\
        e^{i\phi_r}\sin(\frac{\theta_r}{2}) & \cos(\frac{\theta_r}{2})
    \end{pmatrix}.
\end{equation}
The spin and motion are then entangled through a spin-dependent displacement $\hat{D}_{\text{sd}}(\alpha)$ created by applying the ODF beams for a duration $\tau$.  This separates the spin states in phase space as shown in Fig.~\ref{PhaseProto}(a) to form the state
\begin{equation} \label{DispState}
\begin{aligned}
    \ket{\Psi_{3}}&=\hat{D}_{\text{sd}}(\alpha)\hat{R}\left(\frac{\pi}{2},0\right)\ket{\Psi_{1}} \\
    &=\frac{\ket{\uparrow}\hat{D}(\alpha)+\ket{\downarrow}\hat{D}(-\alpha)}{\sqrt{2}}\hat{S}(\xi)\ket{n},
\end{aligned}
\end{equation}
where $\hat{D}(\alpha)\equiv\exp(\alpha \hat{a}^{\dagger}-\alpha^{\ast}\hat{a})$.

A final $\pi/2$-pulse then creates
\begin{equation} \label{FinState}
\begin{aligned}
    \ket{\Psi_{f}}&=\hat{R}\left(\frac{\pi}{2},0\right)\hat{D}_{\text{sd}}(\alpha)\hat{R}\left(\frac{\pi}{2},0\right)\ket{\Psi_{1}} \\
    &=\frac{1}{2}\ket{\uparrow}\left[\hat{D}(\alpha)-\hat{D}(-\alpha)\right]\hat{S}(\xi)\ket{n}\\
    &+\frac{1}{2}\ket{\downarrow}\left[\hat{D}(\alpha)+\hat{D}(-\alpha)\right]\hat{S}(\xi)\ket{n},
\end{aligned}
\end{equation}
and maps this entangled state into a population imbalance in the $\ket{\uparrow}$ and $\ket{\downarrow}$ spin states.  The probability of measuring $\ket{\uparrow}$ for state $\ket{\Psi_{f}}$ is dependent on the overlap between the displaced states
\begin{equation} \label{Pup}
    P^{(n)}_{\uparrow}=\frac{1}{2}+\frac{1}{2}\bra{\Psi_{f}}\hat{\sigma}^z\ket{\Psi_{f}}e^{-\Gamma\tau},
\end{equation}
where we have included a spin decoherence rate $\Gamma$ due to off-resonant light scatter from the ODF laser beams.  Taking a Boltzmann-weighted thermal average over all Fock states, we obtain 
\begin{equation} \label{BrightFrac}
    P_{\uparrow}=\frac{1}{2}-\frac{1}{2}e^{-\Gamma\tau}\exp\left[-\frac{1}{4N}|f\tau|^{2}\left(2\bar{n}_{z}+1\right)\chi(r,\Delta\phi)\right],
\end{equation}
where 
\begin{equation} \label{Squeezed}
    \chi(r,\Delta\phi) = e^{2r}\left[1+\cos(\Delta\phi)\right]+e^{-2r}\left[1-\cos(\Delta\phi)\right]
\end{equation}
contains the dependence on the parametric drive parameters.  Equation~\ref{BrightFrac} reduces to Eq. (A5) of Ref.~\citenum{SawyerApp} when no parametric drive is applied ($r=0$).  The population in $\ket{\uparrow}$ is measured by pulsing on the parallel Doppler cooling beam and counting the resulting fluorescence.  This is referred to as the bright fraction throughout this paper.

Figure~\ref{PhaseProto}(b) shows the resulting bright fraction versus the relative phase $\Delta\phi$ for a crystal of $N=86\pm10$ ions.  Each data point (black circles) is an average over $50$ experimental trials.  Using Eqs.~\ref{BrightFrac} and~\ref{Squeezed} to fit these data, we extract $r=1.25\pm0.2$, where the uncertainty in $r$ results from the scatter of the data and the uncertainty in the temperature $\bar{n}_{z}=0.38\pm0.2$.  This agrees well with the predicted $r = 1.3\pm0.2$ from the applied parametric drive voltage and duration.  The green solid curve shows this theory fit, where the shaded region represents the uncertainty in the measured $\bar{n}_z$.  When the spin-dependent displacement is along the squeezed axis $(\Delta\phi = 0,2\pi)$, the overlap between the displaced states is small and the bright fraction is high.  In contrast, when $\hat{D}_{\text{sd}}(\alpha)$ is orthogonal to the squeezed axis, there is strong overlap between the states resulting in a low bright fraction.  The motional uncertainty of the fitted squeezed state has been attenuated by $5.4 \pm 0.9$ dB (10.8 dB in variance) below the ground-state motional uncertainty.  Higher squeezing might be possible, as will be discussed in Sec.~\ref{ConProto}, but higher squeezing will also be more sensitive to noise in $\Delta\phi$, providing a trade-off in the utility of employing higher squeezing.

\subsection{Improved Displacement Sensitivity}

One application of this stroboscopic squeezing protocol is improving the experimental sensitivity for detecting small displacements.  Experiments conducted with a single ion~\cite{BurdPara} demonstrated an enhancement of $17.2\pm0.3$\,dB to the displacement sensitivity with parametric amplification.  On the large 2D crystals confined in this device, a displacement sensitivity of $8.8\pm0.4$\,dB below the standard quantum limit has recently been achieved~\cite{GilmoreScience}.  In this section, we explore theoretically improving this sensitivity through the addition of parametric amplification in a crystal of ions.

\begin{figure}
\includegraphics[width=0.46\textwidth]{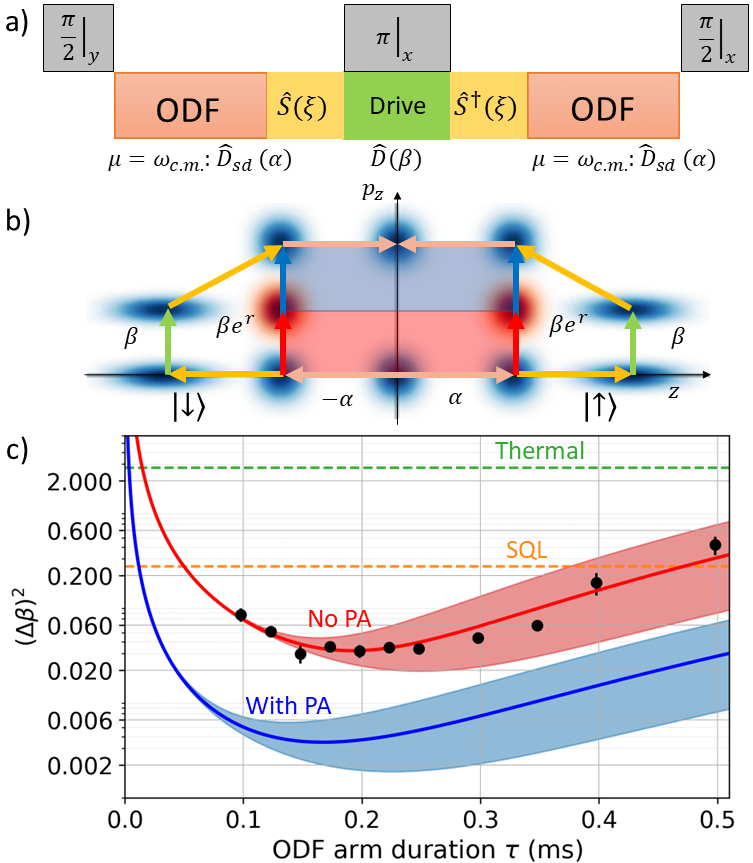}%
\caption{(a) and (b) After Doppler cooling and initializing the ions in the spin up state, a many-body spin echo enables detection of small displacements of the ion crystal caused by weak electric fields resonant with the c.m.\ motion (green block and arrows).  The addition of the squeezing and antisqueezing operations (yellow blocks and arrows) amplifies this spin-independent displacement $\beta$ (red and green arrows) by $e^r$ (effective displacement, blue arrows). The amplified effective displacement $\beta e^r$ (blue arrows) encloses a larger phase-space area compared to no squeezing, as shown by the red area without squeezing and the additional enclosed blue area from the effective amplified displacement. This larger enclosed phase is predicted to improve the sensitivity to small displacements.  (c) Previous experimental (black points) and theoretical (red curve) work \cite{GilmoreScience} without parametric amplification showed the displacement sensitivity to be limited by $40$\,Hz frequency fluctuations of the c.m.\ mode. Confidence bands represent c.m.\ frequency fluctuations between 20 and 60 Hz.  The addition of motional squeezing will amplify this frequency noise, but displacement sensitivity nearly $19$\,dB below the SQL (orange line) is still predicted (blue curve), assuming the duration of the  squeezing and displacement are short compared to $\tau$.}
\label{DispFig}
\end{figure}

Figure~\ref{DispFig}(a) shows the proposed experimental sequence for improved displacement sensing.  When no parametric amplification is applied, this protocol is identical to the sequence used in previous sensing work~\cite{GilmoreScience}.  The c.m.\ motion is entangled with the collective spin via a many-body echo, enabling detection of a small displacement through the resulting spin rotation while avoiding quantum back-action and thermal noise.  By squeezing the motional mode before and antisqueezing after the applied displacement as shown in the phase-space diagram of Fig.~\ref{DispFig}(b), the displacement is amplified, resulting in a larger geometric phase, and ideally improving the detection sensitivity.  The amplification obtained with the squeeze-displace-antisqueeze sequence is given by the identity
\begin{equation} \label{SqIdent}
    \hat{D}_{\text{sd}}(\beta_{f})\equiv\hat{S}^{\dagger}(\xi)\hat{D}_{\text{sd}}(\beta_{i})\hat{S}(\xi),
\end{equation}
where the initial displacement $\beta_{i}$ is amplified by $G=\exp(r)$ to $\beta_{f}=G\beta_{i}$ when the displacement is along the direction of maximum parametric amplification.  Ideally, this would improve the displacement sensitivity $(\Delta\beta)^{2}$, the variance with which $\beta_i$ can be determined in a single measurement, by $\exp(2r)$.  

In Ref.\citenum{GilmoreScience}, experimental measurements and theory showed that the displacement sensitivity of the protocol in Fig. \ref{DispFig} without parametric amplification was limited by $\sigma/=40$\,Hz frequency fluctuations of the c.m.\ mode. The analysis assumed the c.m.\ mode frequency was constant during a single experiment but exhibited Gaussian fluctuations with standard deviation $\sigma$ from one experiment to the next. Expanding the theoretical treatment of Ref.\citenum{GilmoreScience} to include both c.m. frequency fluctuations and parametric amplification, we generalize Eq. (S48) of the Supplementary Information of Ref.\citenum{GilmoreScience} for the sensitivity $(\Delta\beta)^2$  of the protocol shown in Fig.~\ref{DispFig}(a).  We obtain

\begin{equation} \label{DispSens}
\begin{aligned}
    \left(\Delta\beta\right)^{2}&\approx\frac{e^{2\Gamma\tau}}{4f^{2}\tau^{2}e^{2r}}\Bigg[1+\frac{\sigma^{2}\tau^{2}}{3}\\
    &+\frac{\sigma^{2}}{g^{2}}\left(r-\frac{1-e^{-2r}}{2}\right)+\frac{\sigma^{2}\tau}{g}\frac{1-e^{-2r}}{2}\Bigg]\\
    &+\frac{\sigma^{2}\tau^{2}}{2e^{2r}}\left(1+\frac{\sinh(r)e^{r}}{g\tau}\right)^{2}\left(\bar{n}_{z}+\frac{1}{2}\right)\\
    &+\frac{f^{2}\sigma^{2}\tau^{4}}{9e^{2r}},
\end{aligned}
\end{equation}
where the duration of the squeeze and excitation of the spin-independent displacement are assumed to be short compared to $\tau$. Here $\tau$ is the duration of an ODF pulse in a single arm of the spin-echo protocol of Fig.~\ref{DispFig}. When no squeezing is applied $(r=0)$, Eq.~\eqref{DispSens} recovers the prior displacement sensitivity results.  The $\exp(2r)$ term in the denominator expresses the ideal enhancement to the sensitivity if the squeezing only amplified the displacement.  However, the squeezing also increases the noise from frequency fluctuations, which reduces the overall gain. For a more detailed discussion and physical interpretation of the individual terms of Eq.~\eqref{DispSens}, see the Supplementary Information of Ref.\citenum{GilmoreScience}.

In Fig.~\ref{DispFig}(c), we plot the prior experimental data (black circles) and theory (red curve) of the displacement sensitivity without parametric amplification~\cite{GilmoreScience}.  Assuming the same experimental parameters as this previous experiment and the addition of the demonstrated $r=1.25$ of motional squeezing, Eq.~\eqref{DispSens} predicts nearly 10-dB enhancement to our displacement sensitivity (blue curve in Fig.~\ref{DispFig}(c)).  With the assumption that the duration of the squeezing and displacement are short compared to the duration of the spin-dependent force, the quantum enhancements from the many-body echo and the squeezing approximately add, resulting in a displacement sensitivity nearly $19$\,dB below the SQL.  Further improvements would be possible by reducing the temperature of the ion crystal with EIT cooling, increased squeezing, and reduced frequency fluctuations of the c.m.\ mode.

\section{Continuous Protocol}
\label{ConProto}
\subsection{Measuring $g$ From Shifts in the Decoupling Points} 

\begin{figure}
\includegraphics[width=0.46\textwidth]{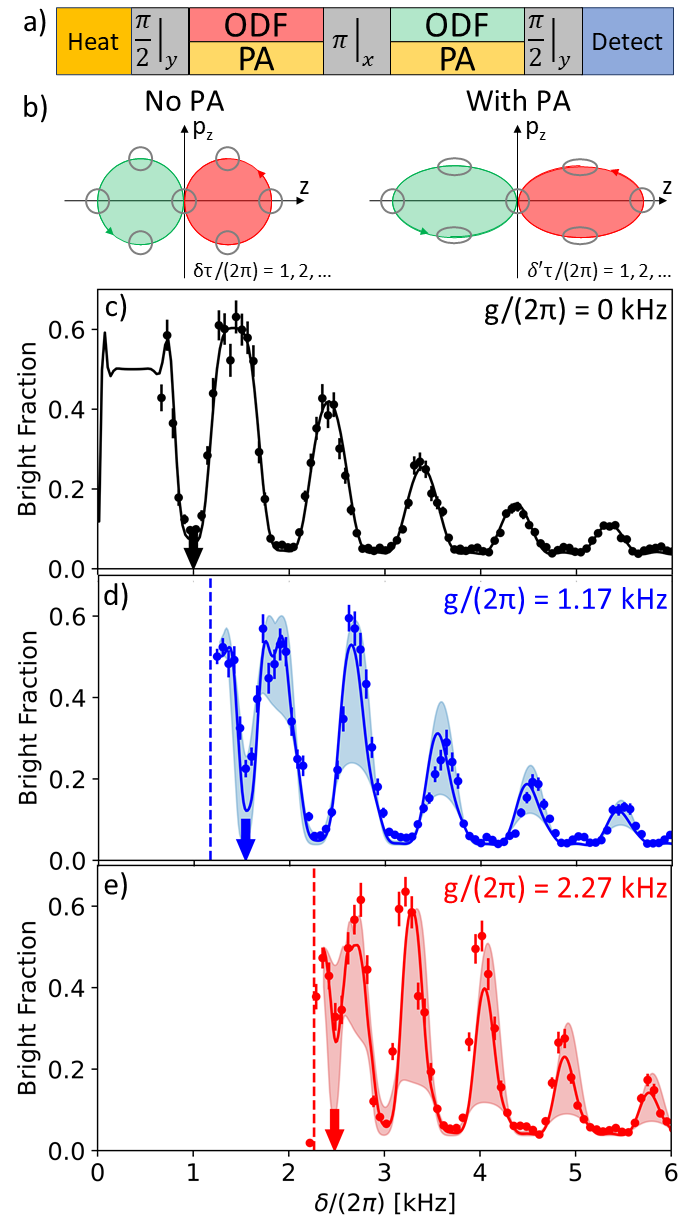}%
\caption{(a) Pulse sequence implemented with the continuous squeezing protocol to measure the parametric coupling strength $g$.  The c.m.\ mode is initially heated to increase the contrast due to spin-motion entanglement.  A spin echo sequence is then used to couple the spin and c.m.\ motion of the ions with the ODF and parametric drive applied simultaneously in each arm for a duration $\tau$.  (b) Phase-space trajectories for the state $\ket{\uparrow}^{\otimes N}$ during the first (red) and the second (green) arm of the pulse sequence. Without parametric amplification, the spin and motion are decoupled in each arm (closed loop in phase space) when $2\pi/\delta$ is an integer multiple of $\tau$.  Parametric amplification squeezes and antisqueezes the motion along this trajectory (gray circles) and shifts the decoupling points to higher frequencies. A new effective detuning $\delta'=\sqrt{\delta^2-g^2}$ sets the condition for closing loops in phase space when $2\pi/\delta'$ is an integer multiple of $\tau$.    (c) Scan over several decoupling points without parametric amplification.  The curve is a theory fit used to extract the elevated $\bar{n}_{z}$ and coherent excitation $\beta$ resulting from the heating pulse. (d) and (e) With parametric amplification the decoupling points (black, blue, and red arrows) are shifted to higher frequency as predicted by theory (curves).  The shaded region reflects the uncertainty in $\Delta\phi_c$ within $[0,2\pi]$. The theory curve then assumes an intermediate value of $\Delta\phi_c=\pi/2$, and all curves are averaged over 40-Hz frequency fluctuations of the c.m.\ mode. Below $\delta=g$ (dashed lines) is not scanned as the ions are rapidly heated.}
\label{DecoupShift}
\end{figure}

Figure~\ref{DecoupShift}(a) shows the continuous squeezing protocol, where the spin-dependent force and the parametric amplification are applied simultaneously \cite{GePara,BurdPara}.  In this protocol, the spin-dependent force is detuned from resonance with the c.m.\ motion by $\delta$.  The parametric amplification is applied at twice the frequency of the spin-dependent force. The total Hamiltonian of the system is now given by
\begin{equation} 
   \begin{split}
    \hat{H}_{\mathrm{T}}&=\frac{\hbar f}{2\sqrt{N}}\left(\hat{a}e^{-i\phi_{\text{ODF}}}+\hat{a}^{\dagger}e^{i\phi_{\text{ODF}}}\right)\sum_{i}^{N}\hat{\sigma}^z_i\\
    &-\hbar\delta\hat{a}^{\dagger}\hat{a} +i\hbar\frac{g}{2}(\hat{a}^{2}e^{-i\theta}-\hat{a}^{\dagger2}e^{i\theta}).
\end{split}
\label{Htotal}
\end{equation}
Under the condition $0<g<\delta$, the total
Hamiltonian can be written in a simple form as
\begin{equation}
\label{Htotal_simple}
    \hat{H}_{\mathrm{T}}=\frac{\hbar }{2\sqrt{N}}\left(f^{\prime\ast}\hat{b}e^{-i\phi_{\text{ODF}}}+f^{\prime}\hat{b}^{\dagger}e^{i\phi_{\text{ODF}}}\right)\sum_{i}^{N}\hat{\sigma}^z_i
    -\hbar\delta^{\prime}\hat{b}^{\dagger}\hat{b},
\end{equation}
by using a Bogoliubov transformation $\hat{b}=\cosh r\hat{a}+ie^{i\theta}\sinh r\hat{a}^{\dagger} $ with $r=\frac{1}{4}\ln (\frac{\delta+g}{\delta-g})$. Here the new effective detuning is $\delta'\equiv\sqrt{\delta^2-g^2}$  and the rescaled strength of the ODF interaction is $f^{\prime}\equiv f \left(\cosh r+e^{i\Delta\phi_c}\sinh r\right)$. The relative phase between the parametric drive and the ODF in the continuous protocol is defined as $\Delta\phi_c=\theta-2\phi_{\text{ODF}}-\pi/2$, so $\Delta\phi_c = 0$ results in the largest amplified effective force $f^{\prime}$.  We note that this different definition gives rise to a $\pi/2$ phase shift relative to the definition of $\Delta\phi$ in Eq.~\eqref{stroboscopic deltaphi} for the stroboscopic protocol. The state evolution under the total Hamiltonian Eq.~\eqref{Htotal_simple} is provided in Appendix~\ref{AppenA}.

The continuous squeezing protocol provides an alternative method for measuring the parametric coupling strength $g$ that is insensitive to shot-to-shot noise in $\Delta\phi_c$. With no parametric amplification, circular trajectories in phase space are driven.  These trajectories are closed when $\delta$ is a multiple of $2\pi/\tau$, which results in a decoupling of the spin and motion of the ions [Fig.~\ref{DecoupShift}(b)].  Here $\tau$ is the duration of the ODF application in each arm of the spin-echo sequence. The area enclosed in the phase-space loop is equal to the acquired geometric phase.  With the parametric drive applied at twice the frequency of the spin-dependent force $\mu$, the circular trajectories are distorted into ellipses as the c.m.\ motion is squeezed and antisqueezed along the path. In addition, the frequency of the first decoupling point \cite{GePara} shifts to
\begin{equation} \label{DecShift}
    \delta=\sqrt{(2\pi/\tau)^{2}+g^{2}}.
\end{equation}
By measuring this frequency shift when the parametric drive is applied, we can extract the parametric coupling strength. The frequency offset $\delta$ required to drive a closed loop in phase space only depends on $\tau$ and $g$ and is independent of the phase of the spin-dependent force relative to the phase of the parametric drive.  Therefore this method of measuring $g$ is insensitive to shot-to-shot fluctuations in the relative phase.

When the spins and motion are decoupled (closed loop in phase space), the measured bright fraction at the end of the sequence given in Fig.~\ref{DecoupShift}(a) will be at a minimum.  To better resolve these decoupling points, the c.m.\ motion is heated.  This improves the resolution by increasing the spin-motion entanglement signal, which depends on the motional temperature \cite{SawyerApp,SawyerModes}.  The heating pulse is applied directly after Doppler cooling.  It is white-noise with a 10-kHz bandwidth centered around $\omega_{z}$.  This noise heats the c.m.\ mode, but the other axial modes remain near the Doppler limit.

The measured bright fraction versus detuning away from the c.m.\ mode is shown in Fig.~\ref{DecoupShift}(c) when no parametric amplification is applied.  As predicted for the ODF duration $\tau=1.0$\,ms, the minimum in the measured bright fraction occurs at multiples of $1/\tau = 1.0$\,kHz.  With only thermal motion, the bright fraction would saturate at $0.5$.  The higher observed bright fraction suggests the addition of a coherent displacement.  This displacement arises from the heating burst not being purely random.  The curve through the data of Fig.~\ref{DecoupShift}(c) is a fit used to extract $\bar{n}_{z}=28.0\pm13$ and the coherent displacement of amplitude $\beta=13.0\pm0.4$. The phase of the coherent displacement is assumed to be random when averaged over many experiments (see Appendix~\ref{AppenA}).  

Figures~\ref{DecoupShift}(d) and \ref{DecoupShift}(e) show equivalent scans, but with the addition of different strengths of parametric amplification.  As the parametric drive voltage is increased, the decoupling points shift to higher frequency as predicted.  From this frequency shift, we extract the parametric coupling strength by solving for $g$ in Eq.~\eqref{DecShift}.  Experimentally, we find that for $\delta<g$ the ions are heated significantly making recovery with Doppler cooling difficult.

The curves of Fig.~\ref{DecoupShift}(d) and \ref{DecoupShift}(e) are theory assuming the $\bar{n}_{z}$ and $\beta$ obtained from the fit of Fig.~\ref{DecoupShift}(c) and $g$ determined from the shift in the first decoupling frequency.  The shaded region reflects the uncertainty in the relative phase between the ODF and parametric drive.  For these experiments, we focus on the shift in the decoupling points, which are insensitive to the relative phase, so the relative phase at the ions is not measured prior to each experiment.  Including this uncertainty, the theory confidence intervals largely encompass the measured values.  The observed increase in the bright fraction at the first decoupling frequency results from spin squeezing (see Appendix~\ref{AppenA} for a detailed theory) and a convolution of the narrow dip in the bright fraction at the decoupling point with the measured 40-Hz frequency fluctuations of the c.m.\ mode.

\begin{figure}
\includegraphics[width=0.46\textwidth]{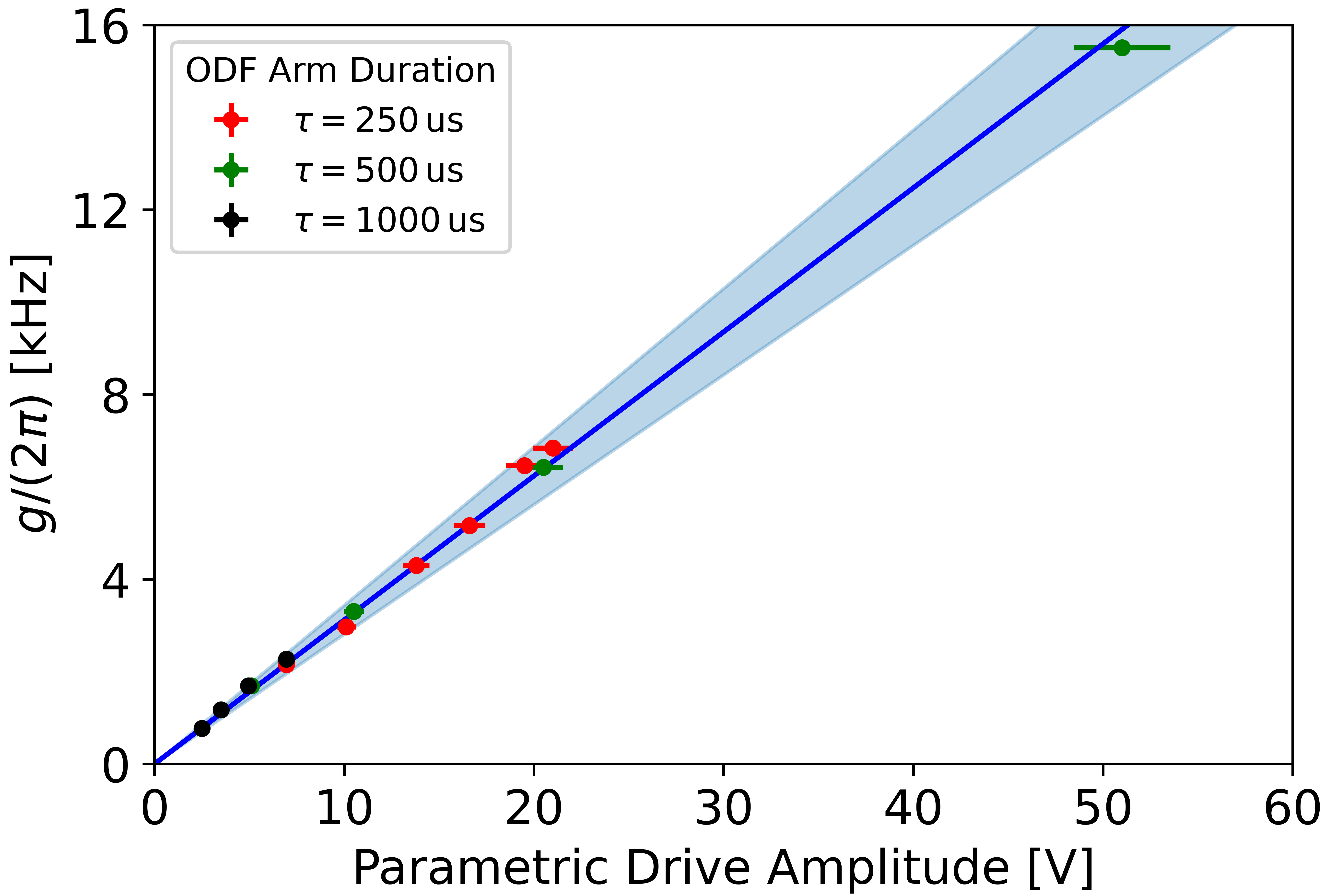}%
\caption{From the shift in the decoupling frequency, the parametric coupling strength $g$ is calculated for a range of drive voltages and ODF arm durations.  As predicted by theory (blue line), $g$ increases linearly with the amplitude of the parametric drive reaching $g/=15.5$\,kHz at $51$~V.  The $5\%$ horizontal error bars of the experimental data represent the uncertainty in the applied voltage and the vertical error bars (smaller than points) reflect the frequency step size of the scans over the decoupling points.  To calculate $g$ from the applied voltages, the trap potential is numerically modeled.  The shaded region is the $10\%$ uncertainty in that model.}
\label{ParaStrength}
\end{figure}

Figure~\ref{ParaStrength} plots the measured $g$ versus the applied parametric drive voltage $V_p$ for three different values of $\tau$.  The coupling strength increases linearly with the voltage.  We see no saturation in $g/$ up to the highest measured value of $15.5$\,kHz at $51.0$\,V, which suggests that even higher parametric coupling strengths may be achievable.  The blue line of Fig.~\ref{ParaStrength} is the predicted coupling strength from the applied voltage, which is in excellent agreement with the measurements.

\subsection{Spin Squeezing Limited by Frequency Fluctuations}
As an application of the continuous squeezing protocol, we investigate how the increased spin-motion coupling will improve quantum spin squeezing in the presence of decoherence due to off-resonant light scatter.  At the decoupling points, the spin-state evolution under the application of Eq.~\eqref{HODF} simulates the Ising model,
\begin{equation} \label{HIsing}
    \hat{H}_{\mathrm{I}}=\frac{1}{N}\sum_{i<j}J_{ij}\hat{\sigma}^z_i\hat{\sigma}^z_j,
\end{equation}
where the spin-spin interaction $J_{ij}\approx\bar{J}=f^{2}/2\delta$ because the c.m.\ mode is primarily driven in our experimental set up \cite{BrittonQSim,BohnetApp}.  This leads to one-axis twisting and quantum spin squeezing \cite{WinelandSqueeze,KitagawaSqueeze}.  Using the continuous parametric amplification protocol, Eq.~\eqref{HODF} transforms into an identical Hamiltonian with rescaled detunings and modified interaction strengths [Eq.~\eqref{Htotal_simple}]. Under optimal relative phase ($\Delta\phi_c=0)$, the spin-dependent force is amplified, resulting in an increased spin-spin interaction $\bar{J}=f^2/2(\delta-g)$. In Ref.~\citenum{GePara} the predicted improvement in quantum spin squeezing from parametric amplification was explored.  Here we extend this analysis to include frequency fluctuations of the c.m.\ mode, which will limit the potential enhancement.

Quantum spin squeezing is characterized by the Ramsey squeezing parameter $\xi_{R}$, where $\xi_{R}^{2}=1$ for coherent spin states and $\xi_{R}^{2}<1$ for squeezed states \cite{WinelandSqueeze} (see Appendix \ref{AppenB}).  In previous experiments on 2D crystals of approximately 100 ions,  $4.0\pm0.9$\,dB of spin squeezing was measured, fundamentally limited by off-resonant light scatter from the optical dipole force laser beams \cite{BohnetApp}.  In these experiments, $\Gamma/\bar{J} \approx 0.05$ where $\Gamma$ is the single spin decoherence rate due to off-resonant light scatter and $\bar{J}$ is the spin-spin interaction strength obtained with a detuning of $\delta/2\pi=1$ kHz.  Parametric amplification will enhance the spin-spin interaction strength $\bar{J}$ while keeping $\Gamma$ fixed by the laser power, which will enable greater spin squeezing.

\begin{figure}
\includegraphics[width=0.46\textwidth]{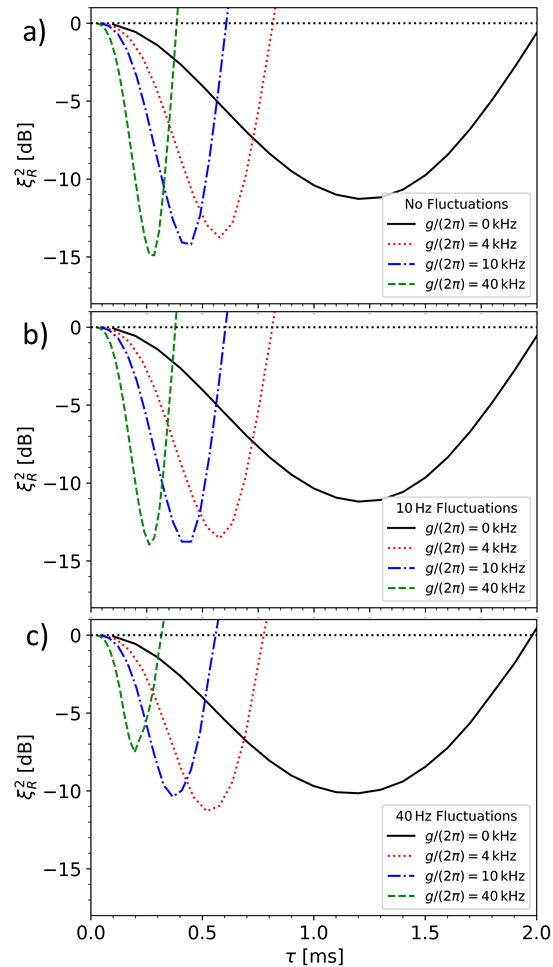}%
\caption{(a) Ramsey squeezing parameter $\xi_{R}$ plotted versus the interaction duration $\tau$ assuming $400$ ions, no frequency fluctuations of the c.m.\ mode, and with a decoherence rate $\Gamma$ due to off-resonant light scatter of $\Gamma = 0.05\bar{J}$ where $\bar{J}$ is the strength of the spin-spin interaction [see Eq.~\eqref{HIsing}] with $\delta/2\pi = 0.83$ kHz and $g=0$. Single-loop gates are assumed where for each interaction duration $\tau$ a detuning $\delta$ given by Eq.~\eqref{DecShift} is assumed.  Larger values of $g$ reduce the effective $\Gamma/\bar{J}$ and produce large spin squeezing.  (b) and (c) Predicted spin squeezing assuming the same parameters as in (a) but now including (b) 10-Hz and (c) 40-Hz frequency fluctuations of the c.m.\ mode.  As the frequency fluctuations are increased, the gain in the quantum spin squeezing with parametric amplification is limited.}
\label{OptSqu}
\end{figure}

Figure~\ref{OptSqu}(a) illustrates the potential enhancement in the spin squeezing with parametric amplification on a crystal containing $400$ ions, cooled to $\bar{n}_{z}=0.5$, and assuming no frequency fluctuations of the c.m.\ mode. Here $\Gamma/\bar{J}=0.05$, where now $\bar{J}$ is the strength of the spin-spin interaction with
$\delta/2\pi = 0.83$ kHz and $g = 0$. When $g=0$, the spin squeezing is limited by spin decoherence from off-resonant light scatter to $11.3$\,dB. We note this larger spin squeezing as compared to Ref.~\citenum{BohnetApp} is due in part to the larger ion number, as well as technical experimental limitations. As $g$ is increased, the ratio $\Gamma/\bar{J}$ is decreased, enabling greater spin squeezing.  The optimal squeezing approaches the spin decoherence-free value of $16.4$\,dB for large $g$.

However, frequency fluctuations of the c.m.\ mode will place a limit on the improvement to spin squeezing from parametric amplification. As the parametric coupling strength increases, the frequency separation between the ODF detuning $\delta$ and $g$ decreases as shown by Eq.~\eqref{DecShift} and Figs.~\ref{DecoupShift}(d) and \ref{DecoupShift}(e).  To realize the maximal optimal spin squeezing free from noise associated with the c.m.\ frequency fluctuations, $\sigma\ll \delta - g$. 

In Figs.~\ref{OptSqu}(b) and \ref{OptSqu}(c) the predicted spin squeezing is plotted as a function of the parametric coupling strength for the same parameters as in Fig.~\ref{OptSqu}(a) but including 10-Hz and 40-Hz frequency fluctuations of the c.m.\ mode.  Spin decoherence from off-resonant light scatter and frequency fluctuations limit the spin squeezing in the absence of parametric amplification.  Figure~\ref{OptSqu}(c) shows a 1.1-dB gain in spin squeezing with a $g/=4$\,kHz and the current 40-Hz frequency fluctuation of the c.m.\ mode.  However, at larger values of $g$, the 40-Hz frequency fluctuations limit further improvements.  In Fig.~\ref{OptSqu}(b), we show that a 2.8-dB gain is possible if the frequency fluctuations are reduced to $10$\,Hz with a $g/2\pi=40$\,kHz.

\section{Conclusion}
\label{Con}
In summary, we have shown experimental results characterizing parametric amplification using two different protocols.  With the stroboscopic protocol, motional squeezing of $5.4\pm0.9$\,dB below the ground-state motion was demonstrated corresponding to a squeezing parameter $r=1.25\pm0.2$. Phase noise between the ODF and parametric drive prevented larger amounts of squeezing from being achieved.  Theory predicts that this level of squeezing will improve the sensitivity of measuring small displacements of large trapped-ion crystals by nearly $10$\,dB.  With the continuous protocol, stronger parametric coupling strengths were demonstrated, even in the presence of shot-to-shot phase noise, by measuring the shift in the frequency of the spin-motion decoupling points.  A linear increase in the measured parametric coupling strength $g$ with the applied parametric drive voltage up to $g/2\pi=15.5\pm0.1$\,kHz was observed. Measured values of $g$ agreed well with the predicted strength from numerical modeling of the trap potentials.  Stronger parametric coupling strengths may be possible with larger applied voltages. An improvement in quantum spin squeezing in the presence of decoherence due to off-resonant light scatter is predicted with the continuous squeezing protocol.  The amount of improvement will be limited by the current 40-Hz frequency fluctuations of the c.m.\ mode.  With reduced frequency fluctuations, further improvements to both the displacement sensitivity and spin squeezing might be possible.
 
\begin{acknowledgments}
We thank Hannah Knaack and Sean Muleady for useful comments and discussions about this manuscript, and David Allcock for the design of the parametric drive circuit.  This work was supported by a collaboration between the U.S. DOE and other agencies.  This material was based upon work supported by the U.S. Department of Energy, Office of Science, NQI Science Research Centers, Quantum Systems Accelerator.  Additional support is acknowledged from AFOSR Grant No. FA9550-201-0019, the DARPA ONISQ program, and NIST. 
\end{acknowledgments}

\appendix
\section{Derivation of the Bright Fraction Under the Continuous Protocol}
\label{AppenA}
 Here we describe the calculation used to generate the theoretical curves of Figs.~\ref{DecoupShift}(c)-\ref{DecoupShift}(e).  Because the continuous protocol of Fig.~\ref{DecoupShift} generates correlations between the ion spins, a single-spin calculation like that described in Sec. III is no longer adequate.
 
 Assume that the spin of the ions is initially in $\ket{\psi}_s=\ket{\uparrow}^{\otimes N}$ and that the c.m.\ motional mode of interest is a thermal coherent state described by a thermal occupation number $\bar{n}_{z}$ and coherent amplitude $\beta$. The initial motional state density matrix is then given by $\hat{\rho}_m=\hat{D}(\beta)\sum_n p_n|n\rangle \langle n|\hat{D}^{\dagger}(\beta)$, where $p_n=\frac{1}{1+\bar{n}_{z}}\left(\frac{\bar{n}_{z}}{1+\bar{n}_{z}}\right)^n$ and $\hat{D}(\beta)=\exp\left(\beta \hat{a}^{\dagger}-\beta^{\ast} \hat{a}\right)$ is the displacement operator with amplitude $\beta$. To derive the effect of the motional thermal coherent state on the bright fraction, it is most convenient to write \cite{SZ} 
\begin{equation}
\hat{\rho}_m=\frac{1}{\pi\bar{n}_{z}}\int e^{-|\gamma|^2/\bar{n}_{z}}\ket{\beta+\gamma}\bra{\beta+\gamma}d^2\gamma.
\end{equation}
Because of the mixture of $\rho_m$ in terms of coherent states $\ket{\bar{\gamma}}$ with $\bar{\gamma}=\beta+\gamma$, we can start with the initial system as
\begin{equation}
\ket{\psi_0}=\ket{\uparrow}^{\otimes N}\ket{\bar{\gamma}}.
\end{equation}
Then we can average out the contribution from different coherent mixtures.

Consider the protocol shown in Fig.~\ref{DecoupShift}(a).  After the $\left.\frac{\pi}{2}\right|_y$ pulse, we have $\ket{\psi_1}=\ket{+}^{\otimes N}\ket{\bar{\gamma}}$, where $\ket{+}=\frac{1}{\sqrt{2}}\left(\ket{\uparrow}+\ket{\downarrow}\right)$. The spin-dependent ODF is then applied simultaneously with the parametric amplification for a duration $\tau$. According to Refs.~[\citenum{GePara}, \citenum{GePRA}], the effective interaction from Eq.~\eqref{Htotal_simple} can be described by 
a product of two unitary operations of the spin-spin interaction and spin-motion coupling in the interaction picture of $ -\hbar\delta^{\prime}\hat{b}^{\dagger}\hat{b}$, which are
\begin{equation}
\hat{U}_{\text{ss}}(t_0,t_1)=\exp\left({i\Phi(t_0,t_1)\sum_{i,j}\hat{\sigma}^z_i\hat{\sigma}^z_j}\right)
\end{equation}
and
\begin{equation}
\hat{U}_{\text{sm}}(t_0,t_1)=\hat{D}_{\text{sd}}[\alpha(t_0,t_1)],
\end{equation}
where 
\begin{equation}
\alpha(t_0,t_1)=\frac{e^{i\phi_{\text{ODF}}}}{2\sqrt{N}}\left[\tilde{\alpha}(t_0,t_1)\cosh r+e^{i\Delta\phi_c} \tilde{\alpha}^{\ast}(t_0,t_1)\sinh r\right]
\end{equation}
and
\begin{equation}
\Phi(t_0,t_1)=-\frac{1}{4N}\left|\frac{f^{\prime}}{\delta^{\prime}}\right|^2\left[(t_1-t_0)\delta^{\prime}-\sin\delta^{\prime}(t_1-t_0)\right].
\end{equation}
Here  $f^{\prime}=f\left(\cosh r+e^{i\Delta\phi_c}\sinh r\right)$ is the enhanced spin-dependent force with parametric drive from Eq.~\eqref{Htotal_simple}, $\tilde{\alpha}(t_0,t_1)=(f^{\prime}/\delta^{\prime})\left(e^{-i\delta^{\prime}t_1}-e^{-i\delta^{\prime}t_0}\right)$ is the displacement of the Bogoliubov mode $\hat b$, and $\alpha(t_0,t_1)$ is the displacement in the original motional mode $\hat a$.

After a simultaneous parametric amplification and ODF pulse for a duration $\tau$, the state is given by
\begin{equation}
\ket{\psi_2}=\hat{U}_{\text{ss}}(0,\tau)\hat{U}_{\text{sm}}(0,\tau)\ket{\psi_1}.
\end{equation}
The $\left.\pi\right|_x$ pulse for a duration $t_{\pi}$ flips the state of the spins. After the second pulse of parametric amplification and ODF, the state is 
\begin{equation}
\begin{aligned}
\ket{\psi_3}&=\hat{U}_{\text{ss}}(\tau+t_{\pi},2\tau+t_{\pi})\hat{U}_{\text{sm}}(\tau+t_{\pi},2\tau+t_{\pi})\hat{R}_X(\pi)\ket{\psi_2}\\
&=\hat{R}_X(\pi)e^{i\Phi_T\sum_{i,j}\hat{\sigma}^z_i\hat{\sigma}^z_j}\hat{D}_{\text{sd}}(\alpha_T)\ket{\psi_1},
\end{aligned}
\end{equation}
where $\hat{R}_X(\pi)=\hat{R}(\pi,\pi/2)^{\otimes N}$ [see Eq.~\eqref{RotationMatrix}] and the expression has been simplified by commutating $\hat{R}_X(\pi)$ with the operators to its right, combining the spin-spin interaction terms by recognizing $\hat{U}_{\text{ss}}(\tau+t_{\pi},2\tau+t_{\pi})=\hat{U}_{\text{ss}}(0,\tau)$, and collecting the spin-spin interaction and the spin-motion coupling terms, respectively. Here the total displacement is
\begin{equation}
\begin{aligned}
\alpha_T&\equiv\alpha(0,\tau)-\alpha(\tau+t_{\pi},2\tau+t_{\pi})\\
&=\frac{e^{i\phi_{\text{ODF}}}}{2\delta^{\prime}\sqrt{N}}\Big[f^{\prime}(e^{-i\delta^{\prime}\tau}-1)(1-e^{-i\delta^{\prime}(\tau+ t_{\pi})})\cosh r\\
&+e^{i\Delta\phi_c}f^{\prime\ast}(e^{i\delta^{\prime}\tau}-1)(1-e^{i\delta^{\prime}(\tau+ t_{\pi})})\sinh r\Big],
\end{aligned}
\end{equation}
and the effective total geometric phase is
\begin{equation}
\begin{aligned}
\Phi_T&\equiv2\Phi(0,\tau)-\text{Im}[\alpha(\tau+t_{\pi},2\tau+t_{\pi})\alpha^{\ast}(0,\tau)]\\
&=\frac{|f^{\prime}|^2}{2\delta^{\prime2}N}\Big\{\sin(\delta^{\prime}\tau)-\delta^{\prime}\tau+[1-\cos(\delta^{\prime}\tau)]\sin\delta^{\prime}(\tau+t_{\pi})\Big\},
\end{aligned}
\end{equation}
which recovers the results in Ref.~\citenum{JordanEIT} when we take $r=0$.
To measure the bright fraction, we rotate $\ket{\psi_3}$ by another $\left.\frac{\pi}{2}\right|_y$ pulse and detect $\hat{\sigma}^z_k$, which gives
\begin{equation}
\begin{aligned}
\braket{\hat{\sigma}^z_k}_{\bar{\gamma}}&=\bra{\psi_3}\hat{R}_Y(\pi/2)^{\dagger}\hat{\sigma}^z_k\hat{R}_Y(\pi/2)\ket{\psi_3}\nonumber\\
&=-\frac{e^{-2|\alpha_T|^2}}{2}(e^{2\alpha_T\bar{\gamma}^{\ast}-2\alpha_T^{\ast}\bar{\gamma}}+\text{c.c.})\cos(4\Phi_T)^{N-1},
\end{aligned}
\end{equation}
where $\hat{R}_Y(\pi/2)=\hat{R}(\pi/2,0)^{\otimes N}$.  Now we sum up the contributions from different components in the initial motional thermal coherent state via the integral
\begin{equation}
\braket{\hat{\sigma}^z_k}=\frac{1}{\pi\bar{n}_{z}}\int e^{-|\gamma|^2/\bar{n}_{z}}\braket{\hat{\sigma}^z_k}_{\bar{\gamma}}d^2\gamma.
\end{equation}
Only the factor $e^{2\alpha_T\bar{\gamma}^{\ast}-2\alpha_T^{\ast}\bar{\gamma}}+\text{c.c.}=e^{2\alpha_T\beta^{\ast}-2\alpha_T^{\ast}\beta}e^{2\alpha_T\gamma^{\ast}-2\alpha_T^{\ast}\gamma}+\text{c.c.}$ will be impacted by the integral. We realize that $\frac{1}{\pi\bar{n}_{z}}\int e^{-|\gamma|^2/\bar{n}_{z}}e^{2\alpha_T\gamma^{\ast}-2\alpha_T^{\ast}\gamma}d^2\gamma=e^{-4|\alpha_T|^2\bar{n}_{z}}$; therefore, we find
\begin{equation}
\label{sigmaz}
\braket{\hat{\sigma}^z_k}=-e^{-2|\alpha_T|^2(2\bar{n}_{z}+1)}\cos(4\theta_{\beta})\cos(4\Phi_T)^{N-1},
\end{equation}
where $\theta_{\beta}=\text{Im}(\beta^{\ast}\alpha_T)$. The coherent displacement $\beta$ in this experiment is caused by the application of noise to heat the ions.  We therefore assume that the phase of this displacement varies from shot to shot, so we average Eq.~\eqref{sigmaz} over a random phase $\theta_c$, giving $\frac{1}{2\pi}\int \cos(4|\alpha_T||\beta|\sin\theta_c)d\theta_c=J_{0}(4|\alpha_T||\beta|)$, where $J_{0}$ is the zeroth Bessel function of the first kind.  The spin decoherence factor $\exp(-2\Gamma\tau)$ can be included independently. The bright fraction of the trapped-ion spins at the end of the pulse sequence is then given by
\begin{equation}
\begin{aligned}
P_{\uparrow}&=\frac{1}{2}\left(1+\braket{\hat{\sigma}^z_k}\right)\nonumber\\
&=\frac{1}{2}-\frac{1}{2}e^{-2|\alpha_T|^2(2\bar{n}_{z}+1)}e^{-2\Gamma\tau}J_{0}(4|\alpha_T||\beta|)\cos(4\Phi_T)^{N-1}.
\end{aligned}
\end{equation}
This expression was used to model the experimental measurements in Fig.~\ref{DecoupShift}.  To include 40-Hz frequency fluctuations of the c.m.\ mode, we averaged this expression assuming Gaussian frequency fluctuations of the mode.

\section{Spin Squeezing with Spin Decoherence}
\label{AppenB}
This appendix discusses the calculation of quantum spin squeezing in the presence of decoherence due to off resonant light scatter and due to frequency fluctuations of the c.m.\ mode.  The calculation was used to generate the theory curves of Fig.~\ref{OptSqu}.

Quantum spin squeezing in a direction rotated by $\psi$ (from the $+z$ direction) about the $x$ axis is defined by
$\xi_{\psi}^2=N(\Delta \hat{S}_{\psi})^2/|\braket{\textbf{S}}|^2$,
where $\hat{S}_{\psi}=\cos(\psi)\hat{S}_z-\sin(\psi)\hat{S}_y$, $(\Delta \hat{S}_{\psi})^2=\braket{ \hat{S}_{\psi}^2}- \braket{\hat{S}_{\psi}}^2$, 
and $\textbf{S}=\frac{1}{2}\sum_i\big(\hat{\sigma}^x_i,\hat{\sigma}^y_i,\hat{\sigma}^z_i\big)$. The Ramsey spin squeezing is obtained by minimizing $(\Delta \hat{S}_{\psi})^2$ over the angle $\psi$,
\begin{equation}
\begin{aligned}
\xi_R^2&=\frac{N \min_{\psi}[(\Delta \hat{S}_{\psi})^2]}{|\braket{\textbf{S}}|^2}\\
&=\frac{N}{2|\braket{\textbf{S}}|^2}\big\{(\Delta \hat{S}_{y})^2+(\Delta \hat{S}_{z})^2\\
&-\sqrt{[(\Delta \hat{S}_{y})^2-(\Delta \hat{S}_{z})^2]^2+4\text{Cov}\big( \hat{S}_{y},  \hat{S}_{z}\big)^2}\big\},
\label{eq:SSangle}
\end{aligned}
\end{equation}
where the optimal angle is 
\begin{equation}
\psi_{\text{opt}}=\frac{1}{2}\arctan\{2\text{Cov}\big( \hat{S}_{y},  \hat{S}_{z}\big)/[(\Delta \hat{S}_{y})^2-(\Delta \hat{S}_{z})^2]\}
\end{equation}
with 
\begin{equation}
\text{Cov}\big( \hat{S}_{y},  \hat{S}_{z}\big)\equiv \frac{1}{2}\braket{\hat{S}_{y}\hat{S}_{z}+\hat{S}_{z}\hat{S}_{y}}-\braket{\hat{S}_{y}}\braket{\hat{S}_{z}}.
\end{equation}
In the presence of dephasing at rate $\Gamma_{\rm el}$ and spontaneous spin flips from $\ket{\uparrow}$ to $\ket{\downarrow}$ ($\ket{\downarrow}$ to $\ket{\uparrow}$) at rate $\Gamma_{\rm ud}$ ($\Gamma_{\rm du}$), spin dynamics due to an Ising interaction in the $\hat{z}$ basis can be modeled by a master equation in the Lindblad form, which can be solved exactly \cite{FossSS}. For completeness, we quote the spin correlation functions in the case of uniform coupling, i.e. $J_{ij}=J$ for all $i$ and $j$,
\begin{align}
\label{eq:spincor}
&\braket{\hat{\sigma}^{+}_i}=\frac{e^{-\Gamma t}}{2}\Phi^{N-1}(J,t)e^{-2|\alpha|^2(2\bar{n}_{z}+1)},\nonumber\\  
&\braket{\hat{\sigma}_i^a\hat{\sigma}_j^b}=\frac{e^{-2\Gamma t}}{4}\Phi^{N-2}((a+b)J,t)e^{-2|\alpha|^2(a+b)^2(2\bar{n}_{z}+1)},\nonumber\\
&\braket{\hat{\sigma}_i^a\hat{\sigma}_j^z}=\frac{e^{-\Gamma t}}{2}\Psi(aJ,t)\Phi^{N-2}(aJ,t)e^{-2|\alpha|^2(2\bar{n}_{z}+1)},
\end{align} 
where $a,b\in \{+,-\}$,
\begin{align}
\Phi(J,t)&=e^{-\left(\Gamma_{\text{ud}}+\Gamma_{\text{du}}\right)t/2}\Big\{\cos\big[t\sqrt{(2i\gamma+2J/N)^2-\Gamma_{\text{ud}}\Gamma_{\text{du}}}\big]\nonumber\\
&+t\frac{\Gamma_{\text{ud}}+\Gamma_{\text{du}}}{2}\text{sinc}[t\sqrt{(2i\gamma+2J/N)^2-\Gamma_{\text{ud}}\Gamma_{\text{du}}}]\Big\},\nonumber\\
\Psi(J,t)&=e^{-\left(\Gamma_{\text{ud}}+\Gamma_{\text{du}}\right)t/2}t\left[i(2i\gamma+2J/N)-2\gamma\right]\nonumber\\
&\times\text{sinc}\big[t\sqrt{(2i\gamma+2J/N)^2-\Gamma_{\text{ud}}\Gamma_{\text{du}}}\big]
\end{align}
and
\begin{align}
\label{eq:effective}
J&=\frac{f^{\prime2}}{\delta^{\prime}}\left(1-\frac{\sin \delta^{\prime}t}{\delta^{\prime}t}\right),\nonumber\\
\alpha&=\frac{1}{\sqrt{N}}\frac{f^{\prime}}{\delta^{\prime}}\left\{\left[\cos( \delta^{\prime}t )-1\right]e^r-i\sin( \delta^{\prime}t )e^{-r}\right\}.
\end{align}
Here $r$, $f^{\prime}$, and $\delta^{\prime}$ are the same as defined in Appendix~\ref{AppenA}.  The above expressions are obtained assuming the initial state is a product state with all spins pointed along the $x$ direction and the motional state is a thermal state. The factor $e^{-2|\alpha|^2(2\bar{n}_{z}+1)}$ describes the effect of spin-motion entanglement when $\alpha\ne0$. Here $\gamma=\left(\Gamma_{\text{ud}}-\Gamma_{\text{du}}\right)/4$, $\Gamma=(\Gamma_{\text{r}}+\Gamma_{\text{el}})/2$, and $\Gamma_{\text{r}}=\Gamma_{\text{ud}}+\Gamma_{\text{du}}$. The numerical results of the plots in Fig.~\ref{OptSqu} (a) are obtained using Eq.~\eqref{eq:SSangle} together with Eqs.~\eqref{eq:spincor}-\eqref{eq:effective}. The numerical results of the plots in Figs.~\ref{OptSqu}(b) and \ref{OptSqu}(c) are obtained by averaging Eq.~\eqref{eq:SSangle} for 4000 randomly Gaussian distributed frequencies at the respective frequency fluctuations. 

\bibliographystyle{apsrev4-1}
\bibliography{bib}

\end{document}